\documentclass[prd,showpacs,twocolumn,
amsmath,amssymb,superscriptaddress,floatfix,nofootinbib]{revtex4-1}
\usepackage[utf8]{inputenc}
\usepackage[sort&compress]{natbib}
\usepackage[normalem]{ulem}
\usepackage{lipsum} 
\usepackage{bm}
\usepackage{times}
\usepackage{amssymb,amsbsy,amsmath,amsfonts}
\usepackage{graphicx}
\usepackage{float}
\usepackage{color}
\usepackage{morefloats}
\usepackage{rotating}
\usepackage{srcltx}
\usepackage{slashed}
\usepackage{subfigure}
\usepackage{multirow}
\usepackage{verbatim}
\usepackage{hyperref}
\usepackage{tabularx}
\usepackage{adjustbox}
\usepackage{xcolor}
\usepackage{nicefrac}
\usepackage{soul}

\setstcolor{red}

\usepackage{hyperref}
\hypersetup{
	colorlinks	=true,
	urlcolor	=blue,
	linkcolor	=blue,
	citecolor	=blue,
	pdftitle	={},
	pdfauthor	={Zhi-Wei Liu, Jun-Xu Lu, Duo-Lun Ge, Li-Sheng Geng},
	pdfsubject	={Quantum interference effects enhanced in $\pi^+p$ femtoscopic correlation functions}
	}

\begin{document}

\title{Quantum interference effects enhanced in $\pi^+p$ femtoscopic correlation functions}

\author{Zhi-Wei Liu}
\affiliation{Institute for Advanced Study in Nuclear Energy \& Safety, College of Physics and Optoelectronic Engineering, Shenzhen University, Shenzhen 518060, China}
\affiliation{Shenzhen Key Laboratory of Nuclear and Radiation Safety, Shenzhen 518060, China}

\author{Jun-Xu Lu}
\affiliation{School of Physics, Beihang University, Beijing 102206, China}

\author{Duo-Lun Ge}
\affiliation{School of Physics, Beihang University, Beijing 102206, China}

\author{Li-Sheng Geng}
\email[Corresponding author: ]{lisheng.geng@buaa.edu.cn}
\affiliation{School of Physics, Beihang University, Beijing 102206, China}
\affiliation{Sino-French Carbon Neutrality Research Center, \'Ecole Centrale de P\'ekin/School of General Engineering, Beihang University, Beijing 100191, China}
\affiliation{Peng Huanwu Collaborative Center for Research and Education, Beihang University, Beijing 100191, China}
\affiliation{Southern Center for Nuclear-Science Theory (SCNT), Institute of Modern Physics, Chinese Academy of Sciences, Huizhou 516000, China}

\begin{abstract}
We present a comprehensive analysis of the $\pi^+p$ femtoscopic correlation functions measured by the ALICE Collaboration in high‑multiplicity $pp$ collisions at $\sqrt{s}=13$ TeV. Using the Koonin-Pratt formula with a Gaussian source and data‑driven $\pi N$ partial‑wave amplitudes, we account for the contributions from $\pi^+p$ scattering and $\Delta(1232)^{++}$‑decay, thereby successfully reproducing the measured data and their transverse‑mass ($m_T$) dependence. The scattering contribution yields a peak near the relative momentum $k\approx140$ MeV/$c$, whereas the decay contribution peaks around $k\approx220$ MeV/$c$. The observed correlation peak results from a weighted sum of the two contributions, with $m_T$-dependent relative weights. We find that the 140 MeV/$c$ peak originates from quantum interference between the incident and scattered waves—a mechanism previously unnoticed in femtoscopic studies. This finding resolves the peak‑shift puzzle in $\pi^+p$ correlations and provides a novel perspective for quantum interference effects in femtoscopy.
\end{abstract}

\maketitle

\emph{Introduction.} Femtoscopy -- the measurement of two-particle momentum correlation functions (CFs) -- has become a powerful tool in nuclear and particle physics~\cite{Wiedemann:1999qn,Lisa:2005dd,Fabbietti:2020bfg,Liu:2024uxn,Zhang:2026tqs}. Initially introduced to probe the space-time evolution of the strongly interacting medium created in heavy-ion collisions~\cite{Goldhaber:1960sf,Fung:1978eq,Voloshin:1997jh}, it has since been widely applied to measure hadron-hadron interactions across light- and heavy-flavor sectors~\cite{STAR:2014dcy,STAR:2015kha,STAR:2025jwe,ALICE:2019gcn,ALICE:2019hdt,ALICE:2020mfd,ALICE:2021cpv,ALICE:2022enj,ALICE:2024bhk,Si:2025eou}, with demonstrated sensitivity to exotic hadron states~\cite{Kamiya:2022thy,Liu:2023uly,Liu:2023wfo,Liu:2024nac,Vidana:2023olz,Ikeno:2025bsx,Khemchandani:2023xup,Li:2024tvo,Yan:2024aap,Liu:2025eqw}. More recently, femtoscopy has been extended to investigate a long-standing question~\cite{Ono:2018vht,Braaten:2024cke,ALICE:2025byl}: the production mechanism of light (anti)nuclei in high-energy hadronic collisions, which is essential not only for our understanding of nucleosynthesis in collider experiments, but also for modeling the composition of ultrahigh-energy cosmic rays~\cite{PierreAuger:2014sui} and for interpreting potential antinuclei signals in indirect dark-matter searches~\cite{ALICE:2022zuz,Serksnyte:2022onw}.

Using the $\pi^\pm$-deuteron CFs measured in $pp$ collisions at $\sqrt{s}=13$ TeV, the ALICE Collaboration has revealed that (anti)deuteron formation by nucleonic fusion follows the strong decay of short-lived resonances~\cite{ALICE:2025byl}. Model-independent evidence for this picture comes from observing the residual correlation of $\pi^\pm p$ pairs stemming from the $\Delta(1232)$ decay in the $\pi^\pm$-deuteron CF~\cite{ALICE:2025aur}. 
The long‑standing question of (anti)deuteron formation in extreme environments thus appears settled; yet a new puzzle immediately emerges. Both the $\pi^+$-deuteron and $\pi^+p$ correlation data exhibit peaks that are markedly shifted to lower momenta, significantly deviating from the position expected for the $\Delta(1232)$. This observation directly challenges the prevailing paradigm that resonant signals in femtoscopy can be universally captured by a Breit-Wigner parametrization~\cite{Kamiya:2019uiw,ALICE:2022mxo,ALICE:2023wjz,ALICE:2023eyl}.

In an effort to explain these anomalous shifts, the ALICE Collaboration introduced temperature‑dependent $\Delta(1232)$ spectral functions, but the extracted temperatures -- around 20 MeV (for $\pi^+$-deuteron) and 25 MeV (for $\pi^+p$)~\cite{ALICE:2025byl,ALICE:2025aur} -- are dramatically lower than the kinetic freeze‑out temperature (above 100 MeV) obtained from blast‑wave fits to hadron spectra in $pp$ collisions~\cite{ALICE:2020nkc}. In an alternative approach, solving relativistic kinetic equations for $\pi$-catalyzed reactions ($\pi NN\rightarrow\pi d$) shows that the shifts can be reproduced only if the $\Delta(1232)$ mass is reduced by roughly 70 MeV/$c^2$~\cite{Zhang:2025tfd}. Lately, incorporating the $P$‑wave $\pi N$ final‑state interaction via the Friedrichs-Lee model $T$‑matrix has revealed that the finite spatial extent of the emission source renders the system sensitive to off‑shell dynamics, shifting the peak and generating a dip on its high‑momentum side. However, the predicted correlation strength still falls short of the measured amplitude, and the expected high‑momentum‑side dip is absent in the data~\cite{Zhang:2026psh}. The challenges encountered in describing the ALICE data call for an explanation, which is essential not only for clarifying the $\pi^+p$ correlation itself and for interpreting the deuteron formation mechanism via $\pi^+$-deuteron femtoscopy, but also for using Femtoscopy as a precision tool in extracting strong interactions among various hadrons.

Another motivation for a detailed study of the $\pi^+p$ CF is that it provides an ideal system for characterizing the particle-emitting source produced in high-energy hadronic collisions. Similar to the case of nucleon-nucleon scattering, the $\pi N$ interaction is tightly constrained by a wealth of high-precision scattering data~\cite{Workman:2012hx}, which allows for a thorough and model-independent understanding of the final-state interaction. Calibrating the source properties via $pp$ and $\pi p$ CFs is therefore an essential step, laying the groundwork for reliable extractions of two-particle interactions from femtoscopic studies of other hadron pairs. Therefore, in this letter, we aim to carry out a comprehensive, model-independent treatment of the $\pi^+p$ final-state interaction, provide a reliable description of the measured CF, and reveal the physical origin of the peak’s evolution with transverse mass ($m_T$).

\emph{Theoretical framework.} To fully account for the final-state interactions in $\pi^+p$ elastic scattering, we employ the partial-wave $T$-matrix as follows:
\begin{align}
T_{LJ}(k,k^\prime)=T_{LJ}^{\rm str}(k,k^\prime)+T_L^{\rm cou}(k,k^\prime),\label{Eq:T-matrix}
\end{align}
where $T_{LJ}^{\rm str}$ and $T_L^{\rm cou}$ denote the strong and Coulomb interaction parts, respectively. Given the abundant $\pi N$ scattering data, refined partial-wave analyses of the strong interaction amplitudes are available. We therefore adopt the current GWU Data Analysis Center solution as our strong interaction input~\cite{Workman:2012hx}. Here, $T_{LJ}^{\rm str}$ includes two additional step functions $\theta(\Lambda_F-k)\cdot\theta(\Lambda_F-k')$ to model off-shell effects of the amplitude~\cite{Molina:2025lzw,Liu:2024nac}. For the Coulomb part, the coordinate space Coulomb force is Fourier-transformed into momentum space~\cite{Torres-Rincon:2023qll,Encarnacion:2024jge}, with relativistic corrections taken into account (see the Supplementary Materials for details).

With the obtained $T$-matrix, the radial part of the scattering wave function reads
\begin{align}
\mathcal{R}_{LJ}(k,r)=\int_0^{|\boldsymbol{k^\prime}|<\Lambda_F}{\rm d}\boldsymbol{k^\prime}~\mathcal{G}(k^\prime)~T_{LJ}(k,k^\prime)~j_L(k^\prime r),\label{Eq:R-WF}
\end{align}
where $j_L$ is the $L$th order spherical Bessel function. The relativistic two-particle propagator is expressed as
\begin{align}
\mathcal{G}(k^\prime)=\frac{2M}{(2\pi)^3}\frac{E_M+E_m}{2E_ME_m}\frac{1}{s-(E_M+E_m)^2+{\rm i}\varepsilon},\label{Eq:G-propagator}
\end{align}
where $\sqrt{s}$ represents the center-of-mass energy, $E_M=\sqrt{M^2+k^{\prime2}}$ and $E_m=\sqrt{m^2+k^{\prime2}}$ are the intermediate-state energies of the proton and the $\pi^+$ meson, respectively, with $M$ and $m$ their masses. In Eq.~\eqref{Eq:R-WF}, the upper integration limit arises from the step function $\theta(\Lambda_F-k')$, while the other step function $\theta(\Lambda_F-k)$ generally does not contribute because the momentum range of interest is smaller than $\Lambda_F$. It was pointed out that the off-shell ambiguity of the strong interaction may lead to changes in the short-distance behavior of the wave function, thereby introducing uncertainty in the calculation of CFs~\cite{Epelbaum:2025aan}. By varying $\Lambda_F$ from 0.6 to 1.4 GeV/$c$, we find that the resulting uncertainty is negligible. Therefore, only the results for $\Lambda_F=1$ GeV/$c$ are shown in the following.

In theory, the femtoscopic CF is usually calculated via the Koonin-Pratt formula~\cite{Koonin:1977fh,Pratt:1990zq}, which requires two essential inputs: the particle-emitting source and the outgoing wave function. The source characterizes the probability distribution of the relative distance $r$ between the two particles produced in $pp$, $pA$, and $AA$ collisions. In the present work, we adopt the widely used Gaussian source with a single parameter $R$, namely, $S_{12}=\exp[-r^2/(4R^2)]/(2\sqrt{\pi}R)^3$. The outgoing wave function encodes not only the final-state interactions of interest but also various quantum effects, such as quantum statistical effects~\cite{Gyulassy:1979yi}, coupled-channel effects~\cite{Kamiya:2019uiw,Haidenbauer:2018jvl}, off-shell effects~\cite{Epelbaum:2025aan}, and the quantum interference effect identified in this work. Here, we explicitly distinguish two contributions to the $\pi^+p$ CF: 1) from the $\pi^+p$ scattering process; 2) from the $\Delta(1232)^{++}\rightarrow\pi^+p$ decay process. The correlation from the first process, $C_{\rm sca}$, can be decomposed into different partial waves as
\begin{align}
C_{\rm sca}(k)=C_{\rm S1/2}+C_{\rm P1/2}+C_{\rm P3/2}+C_{\rm hpw}.\label{Eq:Csca}
\end{align}
The first three terms contain both strong and Coulomb interactions, whose explicit expressions are given by~\cite{Murase:2024ssm,Liu:2026esv,Ge:2025put}
\begin{align}
C_{LJ}(k)=\omega_{LJ}\int_0^\infty{\rm d}\boldsymbol{r}~S_{12}~(2L+1)|j_L+\mathcal{R}_{LJ}|^2,\label{Eq:CscaSP}
\end{align}
where $\omega_{LJ}=(2J+1)/[2(2L+1)]$ is the weight factor for the $\pi^+p$ system. We have confirmed that the strong-interaction contribution from higher partial waves is negligible over the momentum range of interest. Hence, only the Coulomb contributions from higher partial waves are retained, as in Refs.~\cite{Liu:2022nec,Torres-Rincon:2023qll}
\begin{align}
C_{\rm hpw}(k)=\int_0^\infty{\rm d}\boldsymbol{r}~S_{12}~\left(|\Phi^{\rm cou}|^2-\sum_{L=0,1}|\Phi_L^{\rm cou}|^2\right),\label{Eq:Cscahpw}
\end{align}
where $\Phi^{\rm cou}$ and $\Phi_L^{\rm cou}$ denote the full Coulomb wave function and its $L$th order component, respectively. 

The contribution from the second process, $C_{\rm dec}$, reads~\footnote{Since the final‑state $\pi^+p$ pairs from both the scattering and decay channels evolve under the same Hamiltonian, their ``scattered waves'' share the same shape.}
\begin{align}
C_{\rm dec}(k)=\int_0^\infty{\rm d}\boldsymbol{r}~S^{R=0.42}_{12}~(2L+1)|\mathcal{R}_{LJ}|^2,\label{Eq:Cdec}
\end{align}
where $S^{R=0.42}_{12}$ denotes the Gaussian source with $R=0.42$ fm. It is worth noting that $C_{\rm dec}$ differs from $C_{\rm sca}$ (and particularly from its $C_{\rm P3/2}$ component, Eq.~\eqref{Eq:CscaSP}) in three important aspects: (i) The outgoing wave function contains only the ``scattered wave'', with no incident wave $j_L$. As will be shown later, the interference between these two waves is essential to the $\pi^+p$ CF. (ii) Owing to its fixed total angular momentum, $C_{\rm dec}$ is free from the suppression by the weight factor $\omega_{LJ}$. (iii) The source size for $\pi^+p$ pairs from the $\Delta(1232)^{++}$ decay is expected to be significantly smaller than that for pairs produced directly in $pp$ collisions. For a Gaussian source, the distribution $4\pi r^2S_{12}$ peaks at $r=2R$. Treating the pion as pointlike and using the proton radius of $\approx0.84$ fm~\cite{Maisenbacher:2026nau,Bullis:2026ebw}, we therefore set $R=0.42$ fm for this component. We have verified that our conclusions are robust against $\pm50\%$ variations of this source size.

\begin{figure*}[htbp]
  \centering
  \includegraphics[width=1\textwidth]{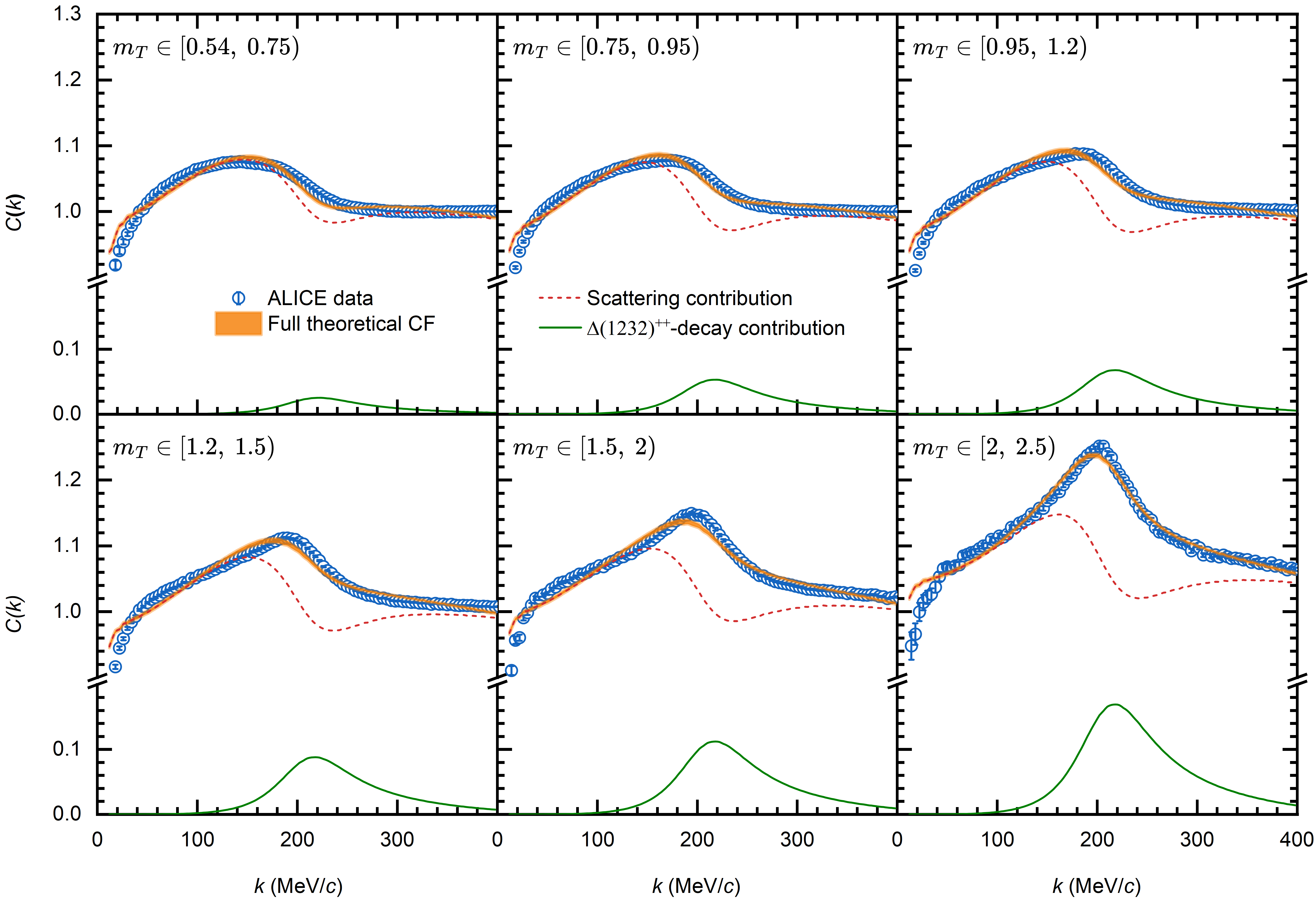}
  \caption{Theoretical $\pi^+p$ CFs as a function of the relative momentum $k$ for six intervals of $m_T$ (in units of GeV/$c^2$) in comparison with the experimental data from $pp$ collisions at $\sqrt{s}=13$ TeV measured by the ALICE Collaboration~\cite{ALICE:2025aur}. The red short‑dashed and green solid lines show the $\pi^+p$ scattering and $\Delta(1232)^{++}$‑decay contributions, respectively, whose sum yields the total theoretical CFs (orange band, $1\sigma$ uncertainty).}\label{Fig:exp vs the}
\end{figure*}

For comparison with the experiment, the following corrections are applied:
\begin{align}
C(k)=(a+bk)[1+\lambda(1-P)(C_{\rm sca}-1)]+\lambda PC_{\rm dec}.\label{Eq:C}
\end{align}
These corrections include a weight factor $P$ giving the separable fraction of the $\Delta(1232)^{++}$-decay contribution (decays after the resonance leaves the source), and a parameter $\lambda$ that incorporates the effects of particle misidentification and feed-down effects. For $C_{\rm sca}$, we also include a background contribution from correlated $\pi^+p$ pairs originating from mini-jets. We parameterize this background with a simple linear form, $a+bk$, to avoid introducing artificial structures. The parameter $\lambda$ has been evaluated by the ALICE Collaboration~\cite{ALICE:2025aur}. Overall, our framework contains 4 free parameters $P$, $R$ (the effective source size corresponding to $C_{\rm sca}$), $a$, and $b$ to describe the measured $\pi^+p$ CF and its transverse-mass dependence.

\emph{Results and discussion.} The ALICE Collaboration has precisely measured $\pi^+p$ CFs in six intervals of $m_T$ in high-multiplicity $pp$ collisions at $\sqrt{s}=13$ TeV. We fit these data using the theoretical framework proposed in this work (the four free parameters are listed in the Supplementary Materials). 
As shown in Fig.~\ref{Fig:exp vs the}, the theoretical results (orange bands) agree well with the experimental data (blue points). In particular, the results correctly reproduce the characteristic shift of the $\pi^+p$ correlation peak toward lower relative momentum $k$ with decreasing $m_T$: for a low-$m_T$ interval [0.54-0.75) GeV/$c^2$ a broad peak appears near $k\approx140$ MeV/$c$, whereas for a high-$m_T$ interval [2-2.5) GeV/$c$ a narrow peak is seen around $k\approx200$ MeV/$c$.

We further decompose the full theoretical results into $(a+bk)[1+\lambda(1-P)(C_{\rm sca}-1)]$ and $\lambda P C_{\rm dec}$ parts, shown as the red short‑dashed and green solid lines, respectively. The scattering part yields a broad peak near $k\approx140$ MeV/$c$ and a dip around $k\approx230$ MeV/$c$, a pattern that resembles the results of Ref.~\cite{Zhang:2026psh}. The decay part exhibits a peak near $k\approx220$ MeV/$c$; notably, this part grows markedly with increasing $m_T$. Altogether, the experimentally observed peak is a weighted sum of the $\pi^+p$ scattering and $\Delta(1232)^{++}$‑decay contributions, and its shift with $m_T$ is a direct consequence of their evolving relative weights.

These findings point to the following physical picture. In $pp$ collisions, a hadron-emitting source is created, from which the pions, protons, and $\Delta$ resonances are emitted. At low $m_T$ (corresponding to low transverse velocities), the $\Delta(1232)^{++}$ resonance has a higher probability of decaying into a $\pi^+p$ pair before escaping the source region; the resulting pair then mixes with directly produced $\pi^+p$ pairs from $pp$ collisions and undergoes final‑state scattering. In contrast, at large $m_T$ (high transverse velocities), the resonance is more likely to leave the source before decaying, leading to a nearly uncontaminated final state of two-body decay. This picture is strongly supported by the extracted parameters $P$ and $R$. As $m_T$ increases, the parameter $P$, which quantifies the fraction of the $\Delta(1232)^{++}$-decay contribution that occurs after the resonance escapes the source, grows steadily, while the effective source radius $R$ decreases. Here $R$ is an effective source size that incorporates the contribution from $\Delta(1232)^{++}$ decays occurring before they escape the source; its reduction with $m_T$ not only reflects the faster escape of resonances from the collision environment but may also be directly linked to the increase in $m_T$ itself, i.e., the so-called transverse‑mass scaling~\cite{ALICE:2020ibs,ALICE:2023sjd}.

\begin{figure}[htbp]
  \centering
  \includegraphics[width=0.42\textwidth]{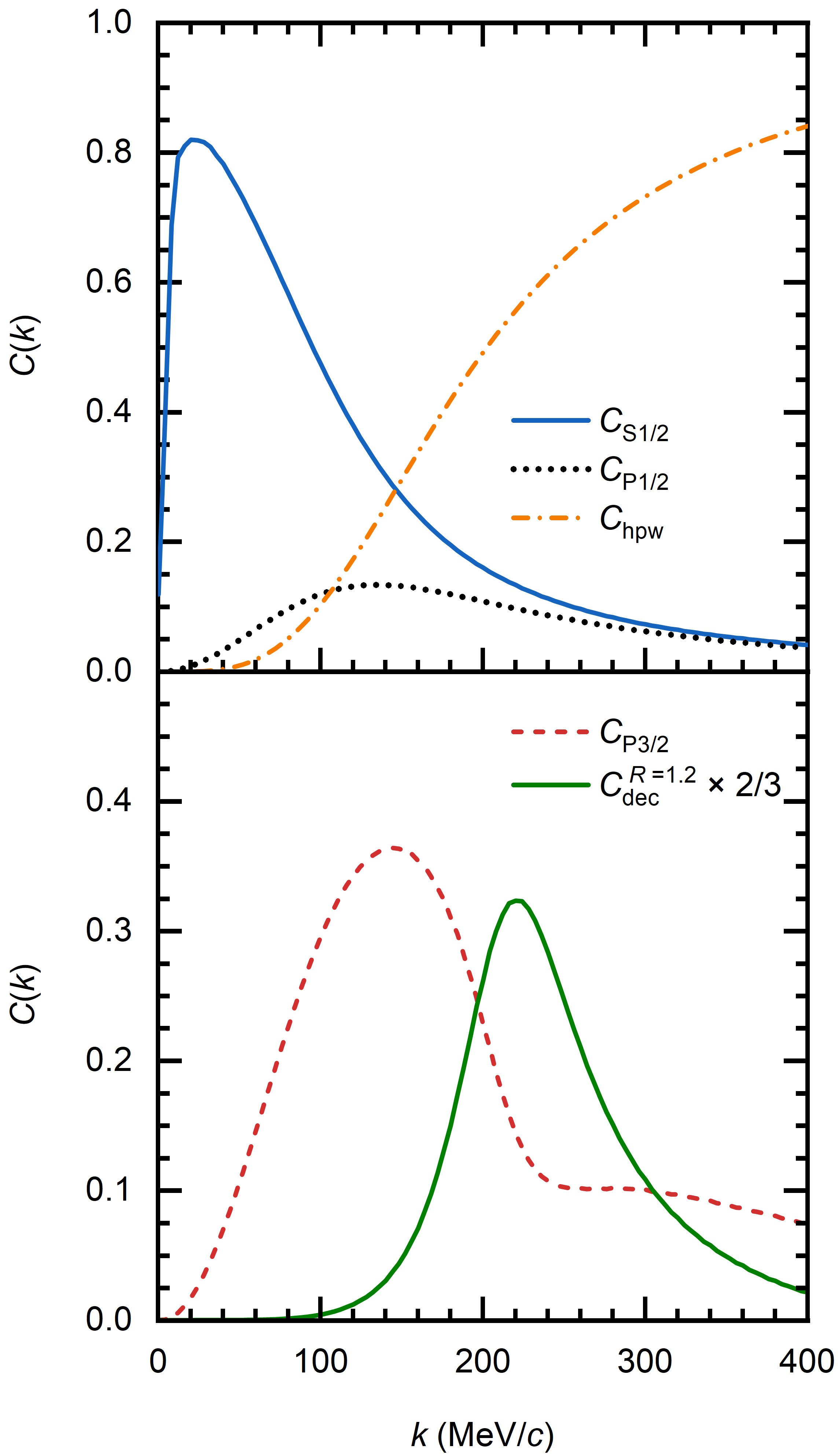}
  \caption{Decomposition of the $\pi^+p$ CF into partial‑wave contributions and the modified $\Delta(1232)^{++}$‑decay channel. The results are calculated with a source size $R=1.2$ fm.}\label{Fig:the ana}
\end{figure}

\begin{figure}[htbp]
  \centering
  \includegraphics[width=0.42\textwidth]{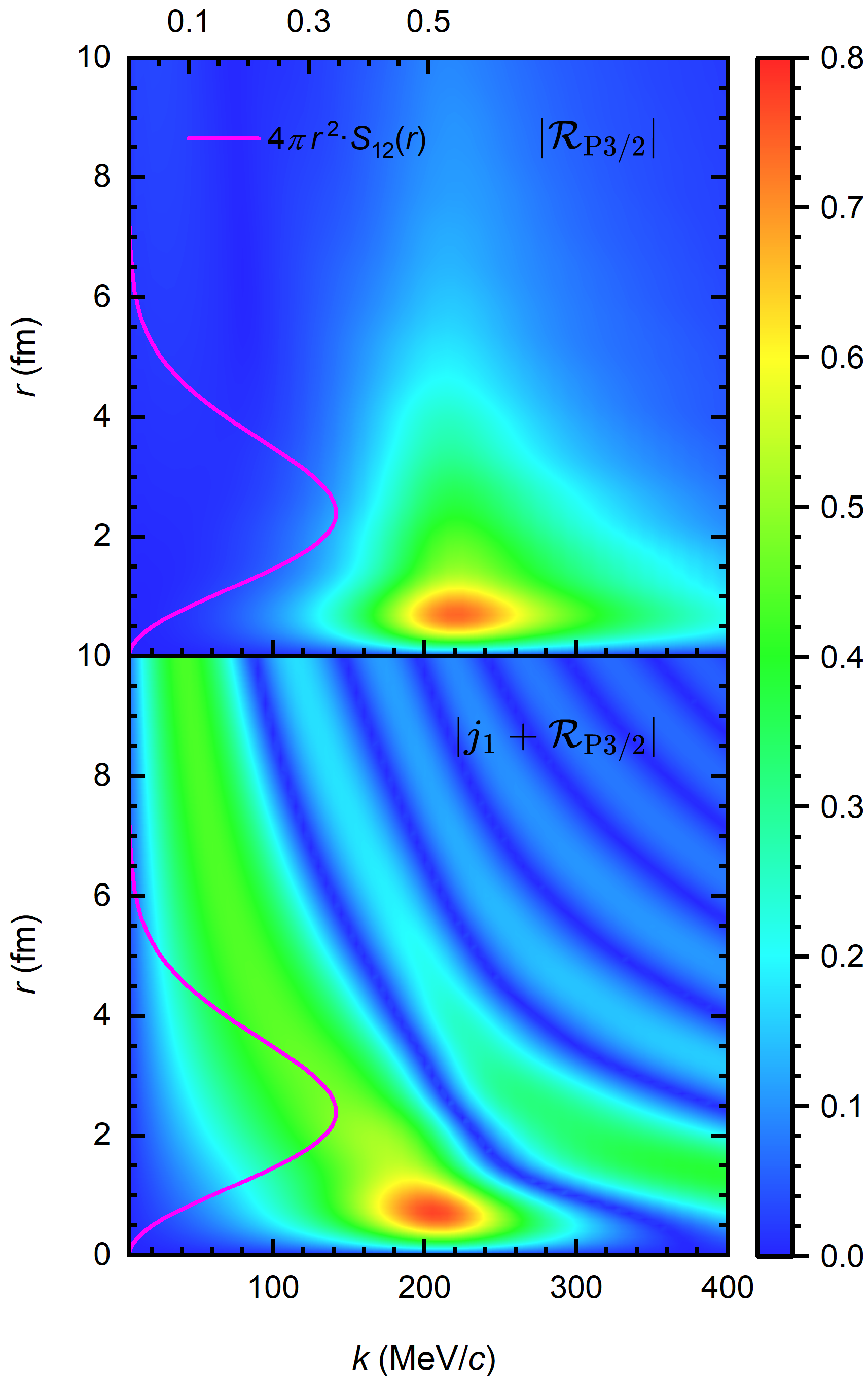}
  \caption{Magnitudes of $|j_1+\mathcal{R}_{\rm P3/2}|$ and $|\mathcal{R}_{\rm P3/2}|$ in the $(k,r)$ phase space. The magenta solid line represents the Gaussian source distribution $4\pi r^2S_{12}$ for $R=1.2$ fm, which is referenced to the tick marks on the upper horizontal axis (in units of fm$^{-1}$).}\label{Fig:wave function}
\end{figure}

To understand why the $\pi^+p$ scattering and $\Delta(1232)^{++}$-decay channels yield peaks at different relative momenta, we compare the contributions from the $S1/2$ (blue solid line), $P1/2$ (black short-dotted line), and $P3/2$ partial waves (red short‑dashed line), the higher partial waves (orange dash-dotted line) in the scattering channel, and the $\Delta(1232)^{++}$-decay channel (green solid line) in Fig.~\ref{Fig:the ana}. For an unbiased comparison, the source size is fixed at $R=1.2$ fm for both the scattering and decay contributions, with the decay contribution additionally multiplied by a weight factor $\omega_{\rm P3/2}=2/3$. It is seen that the $S$‑wave dominates at low momenta, while the higher partial waves dominate at high momenta. The $P$‑wave contributions are most prominent at intermediate momenta; in particular, the $P3/2$ partial wave exhibits a clear peak near $k\approx140$ MeV/$c$, which is the primary origin of the scattering peak. The decay contribution peaks at $k\approx220$ MeV/$c$, consistent with the conventional Breit-Wigner form~\footnote{From the Review of Particle Physics~\cite{ParticleDataGroup:2026}, the $\Delta(1232)^{++}$ Breit-Wigner mass (1230.55 MeV) translates into an $\pi^+p$ relative momentum of 225.83 MeV/$c$.}. Crucially, the only difference between the red short‑dashed and green solid lines is that the former includes the incident wave $j_1$ in the outgoing wave function, whereas the latter describes a pure ``scattered wave''. This inclusion is responsible for the shift of the peak.

We now elucidate how the inclusion of the incident wave $j_1$ shifts the correlation peak toward lower relative momentum. For this purpose, we compare the magnitudes of $|j_1+\mathcal{R}_{\rm P3/2}|$ and $
|\mathcal{R}_{\rm P3/2}|$ in the $(k,r)$ phase space in Fig.~\ref{Fig:wave function}, where the Gaussian source distribution $4\pi r^2S_{12}$ for $R=1.2$ fm is also shown (magenta solid line). The $|\mathcal{R}_{\rm P3/2}|$ amplitude is concentrated mainly in the region $k\gtrsim120$ MeV/$c$ and $r\lesssim5$ fm. The $|j_1+\mathcal{R}_{\rm P3/2}|$ distribution is more complex, with the main feature being a substantial portion of the wave amplitude shifted to lower $k$, which arises from the interference between $j_1$ and $\mathcal{R}_{\rm P3/2}$. For the Gaussian source, the most probable pair separation is $r=2R=2.4$ fm, at which the maxima of $|j_1+\mathcal{R}_{\rm P3/2}|$ and $|\mathcal{R}_{\rm P3/2}|$ lie near $140$ MeV/$c$ and $220$ MeV/$c$, respectively. According to the Koonin-Pratt formula, the overlap integrals of the squared wave functions with the source function then yield correlation peaks at the corresponding momenta.

Before closing this discussion, we emphasize that the large width of the $\Delta(1232)^{++}$ ($\sim112.2$ MeV~\cite{ParticleDataGroup:2026}) is the key reason why quantum interference is so prominent in the $\pi^+p$ CF. This can be contrasted with the $\Lambda(1520)$ resonance in the $K^-p$ system, which has a width of only $\sim16$ MeV~\cite{ParticleDataGroup:2026}, making the corresponding interference effect extremely weak. A re-examination of the $K^-p$ CF confirms this expectation -- the $D$-wave scattering peak shifts by merely 2 MeV/$c$, from 242 MeV/$c$ (without the incident wave) to 240 MeV/$c$ (with the incident wave).

\emph{Summary.} A puzzling feature of the $\pi^+p$ CF -- a pronounced peak that shifts to lower momentum as the transverse mass decreases -- has resisted clear theoretical explanation. We tackle this problem using the Koonin-Pratt formula with a Gaussian emission source and a model-independent treatment of the final‑state interaction, built directly on experimental $\pi N$ scattering data. By cleanly separating the correlation into a $\pi^+p$ scattering contribution and a $\Delta(1232)^{++}$ decay contribution, we successfully described the ALICE measurements, in particular, revealing the physical origin of these peaks.

We find that the scattering part produces a peak near $140$ MeV/$c$, while the decay part peaks around $220$ MeV/$c$. The observed correlation peak is not a single resonance signal but a weighted sum of the scattering and decay contributions, with their relative weights evolving with transverse mass.
The $140$ MeV/$c$ peak in the scattering channel is shown to be a direct consequence of quantum interference between the incident and the scattered waves -- an effect previously unnoticed in femtoscopic studies.

Our findings reveal a fundamental and previously unrecognized role of quantum interference in hadron CFs. This quantum interference mechanism resolves the $\pi^+p$ peak‑shift puzzle and offers a new perspective on quantum interference effects in femtoscopy. The same mechanism may be responsible for the shifts of the $\Delta(1232)$ peaks in $\pi$-deuteron femtoscopic correlation functions, which are being studied and will be reported in a separate work.

\emph{Acknowledgments.} This work is partly supported by the National Science Foundation of China under Grant No. W2543006 and Nos. 12435007 and 1252200936, and the National Key R\&D Program of China under Grant No. 2023YFA1606703. Zhi-Wei Liu acknowledges support from the National Natural Science Foundation of China under Grant No.12405133, and the Shenzhen Science and Technology Program under Grant No.ZDSYS20230626091501002.

\bibliography{pip}

\clearpage
\setcounter{equation}{0}
\setcounter{figure}{0}
\section*{Supplemental Material}
In this Supplemental Material, we present additional details that support the results in the main text.

\section{Strong interaction $T$-matrix}
\begin{figure*}[htbp]
  \centering
  \includegraphics[width=1\textwidth]{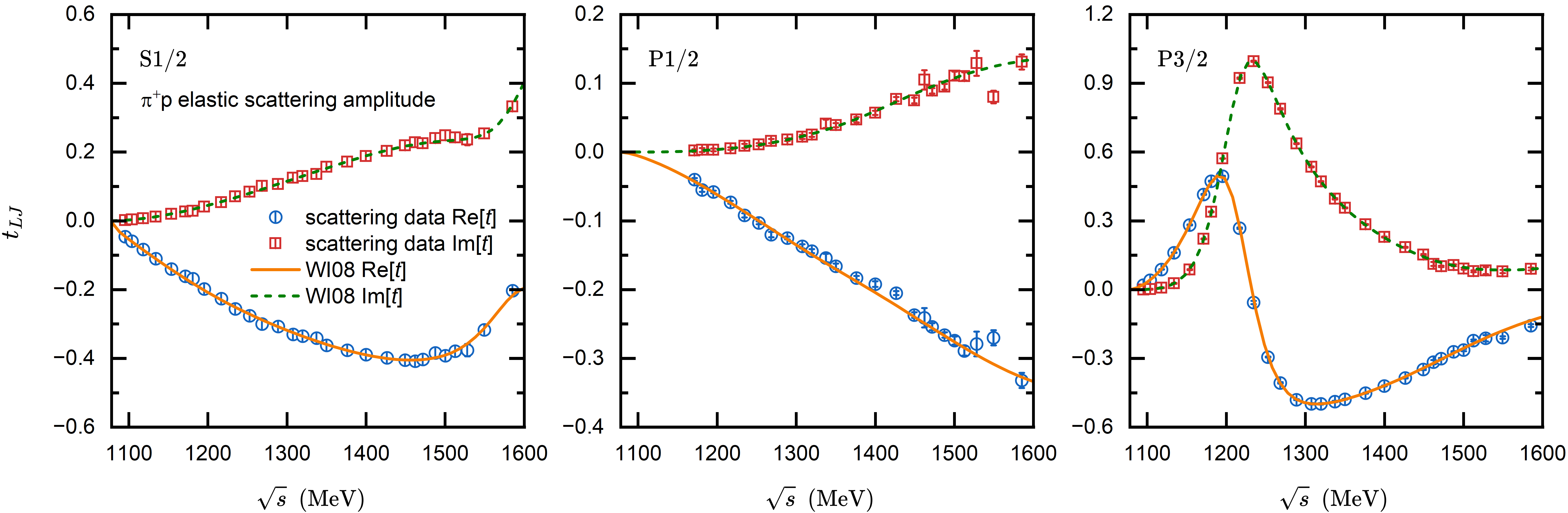}
  \caption{Partial-wave $T$-matrices $t_{LJ}$ from the WI08 solution, compared with $\pi^+p$ elastic scattering data. All results are taken from the SAID program.}\label{Fig:t}
\end{figure*}

For the $\pi N$ elastic scattering, the George Washington University group 
has performed detailed partial-wave analyses, which are widely used to constrain the $\pi N$ interaction in chiral perturbation theory~\cite{Yao:2016vbz,Chen:2012nx} and in the coupled-channel unitary approach~\cite{Inoue:2001ip}. In the present work, we adopt their latest partial‑wave amplitudes, the WI08 solution~\cite{Workman:2012hx}, as our strong-interaction input. Fig.~\ref{Fig:t} shows the $LJ=S1/2$, $P1/2$, and $P3/2$ partial‑wave $T$‑matrices $t_{LJ}$ from this solution. As can be seen, they are in excellent agreement with the experimental data, and the $P3/2$ partial wave clearly supports the existence of the $\Delta(1232)$ resonance. Further analysis reveals that the $S1/2$ and $P1/2$ partial waves exhibit moderate and weak repulsion, respectively. In addition, the contributions of the $D3/2$, $D5/2$, $F5/2$, and $F7/2$ partial waves are found to be negligible.

In our computation of the scattering wave functions, the following substitution is made to ensure a consistent convention~\cite{Inoue:2001ip}:
\begin{align}
T_{LJ}^{\rm str}(p,p^\prime)&=-\frac{1}{\rho}~t_{LJ}~\theta(\Lambda_F-p)\theta(\Lambda_F-p^\prime),
\end{align}
where $\rho=2Mk/(8\pi\sqrt{s})$ is the phase-space factor, $k=\lambda^{1/2}(s,m^2,M^2)/(2\sqrt{s})$ is the center-of-mass momentum, and $\lambda(a,b,c)\equiv a^2+b^2+c^2-2ab-2ac-2bc$ is the K\"allen function. Since the $T$-matrix obtained from the partial‑wave analysis is on‑shell, two additional step functions $\theta(\Lambda_F-k)\cdot\theta(\Lambda_F-k')$ are introduced to model possible off‑shell behavior of the amplitude.

\section{Coulomb interaction $T$-matrix}
For two charged particles, the Coulomb interaction must be taken into account, as it plays a significant role in the low-momentum region. We compute the Coulomb interaction following Refs.~\cite{Torres-Rincon:2023qll,Encarnacion:2024jge}, by first Fourier‑transforming the coordinate‑space Coulomb potential into momentum space
\begin{align}
V^{\rm cou}(|\boldsymbol{p^\prime}-\boldsymbol{p}|)&=\int_0^{|\boldsymbol{r}|<R_c}{\rm d}\boldsymbol{r}~e^{{\rm i}(\boldsymbol{p^\prime}-\boldsymbol{p})\cdot\boldsymbol{r}}\frac{\epsilon\alpha}{r}\nonumber\\
&=\frac{4\pi\epsilon\alpha}{|\boldsymbol{p^\prime}-\boldsymbol{p}|^2}\left[1-\cos(|\boldsymbol{p^\prime}-\boldsymbol{p}|R_c)\right],\label{Eq:couV}
\end{align}
where $\epsilon=1$ corresponds to a repulsive force and $\alpha=1/137.04$ is the fine structure constant. A cutoff $R_c$ is introduced to make the Fourier transform numerically tractable, and at $R_c=60$ fm the results are stable. Subsequently, the Coulomb potential is expanded into partial waves as follows:
\begin{align}
V_L^{\rm cou}(p,p^\prime)=\frac{1}{2}\int_{-1}^1{\rm d}\cos\theta~
V^{\rm cou}(q)P_L(\cos\theta),\label{Eq:couVL}
\end{align}
where $P_L$ is the $L$th order Legendre polynomial and $q^2=p^2+p^{\prime2}-2pp^{\prime}\cos\theta$.

Given the use of a relativistic propagator and the relativistic normalization convention for baryons and mesons in computing the scattering wave functions, the following relativistic correction is applied to the static Coulomb potential:
\begin{align}
T_L^{\rm cou}(p,p^\prime)=&\sqrt{\frac{2E_M(p)2E_m(p)\xi(p)}{2M}}\nonumber\\
&\times\sqrt{\frac{2E_M(p^\prime)2E_m(p^\prime)\xi(p^\prime)}{2M}}V_L^{\rm cou}.\label{Eq:couT}
\end{align}
Here, the first-order Coulomb amplitude in the Bethe-Salpeter equation is replaced by $V_L^{\rm cou}$, which is practically indistinguishable from the fully unitarized one~\cite{Encarnacion:2024jge}. The kinematic factor $\xi$ is given by
\begin{align}
\xi(p)=\frac{\sqrt{s}-E_M(p)-E_m(p)}{k^2/(2\mu)-p^2/(2\mu)},\label{Eq:couxi}
\end{align}
where $\mu=Mm/(M+m)$ is the reduced mass.

\section{Fitting parameters}
The fitting parameters used in Eq.~\eqref{Eq:C} in the main text are listed in Table~\ref{Tab:parameter}.
\begin{table*}[htbp]
\centering
\setlength{\tabcolsep}{16pt}
\caption{Transverse mass $m_T$ intervals and parameter $\lambda$ from the ALICE measurements~\cite{ALICE:2025aur}, together with the fit parameters $P$ (weight factor), $R$ (effective source size), $a$ (background parameter), and $b$ (background parameter) obtained in the present analysis.}\label{Tab:parameter}
\begin{tabular}{ccccccc}
\hline\hline
No. & $m_T$~(GeV$/c^2$)~\cite{ALICE:2025aur} & $\lambda$~\cite{ALICE:2025aur} & $P$  & $R$~(fm)  & $a$  & $b$~($10^{-4}$ MeV$^{-1}$)  \\ 
\hline
I & $[0.54, 0.75)$ & $0.677$ &0.03$\pm$0.00  & 1.32$\pm$0.02  & 1.08$\pm$0.00  & -2.20$\pm$0.08  \\ 
II & $[0.75, 0.95)$ & $0.687$ &0.06$\pm$0.00  & 1.24$\pm$0.02  & 1.06$\pm$0.00  & -1.87$\pm$0.08  \\ 
III & $[0.95, 1.20)$ & $0.689$ &0.08$\pm$0.00  & 1.21$\pm$0.02  & 1.06$\pm$0.00  & -1.83$\pm$0.08  \\ 
IV & $[1.20, 1.50)$ & $0.688$ &0.11$\pm$0.00  & 1.18$\pm$0.02  & 1.07$\pm$0.00  & -1.85$\pm$0.08  \\ 
V & $[1.50, 2.00)$ & $0.661$ &0.14$\pm$0.00  & 1.15$\pm$0.02  & 1.08$\pm$0.00  & -1.87$\pm$0.08  \\ 
VI & $[2.00, 2.50)$ & $0.616$ &0.23$\pm$0.00  & 1.02$\pm$0.01  & 1.13$\pm$0.00  & -1.91$\pm$0.09  \\ \hline\hline
\end{tabular}
\end{table*}

\section{Influence of strong interaction}
\begin{figure}[htbp]
  \centering
  \includegraphics[width=0.42\textwidth]{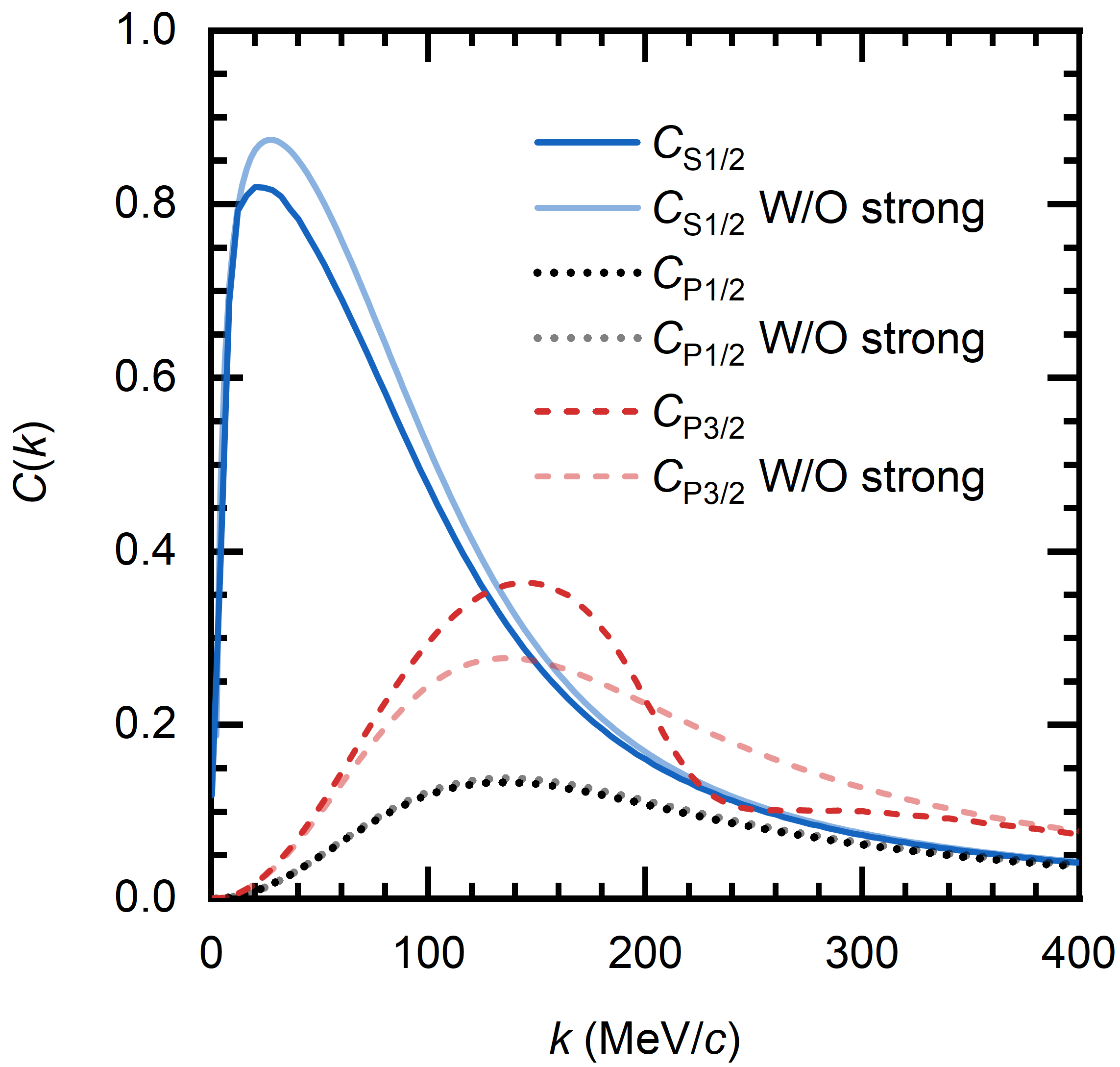}
  \caption{Influence of $S$- and $P$-waves strong interaction in $\pi^+p$ CF. The results are calculated with a source size $R=1.2$ fm.}\label{Fig:the ana without strong}
\end{figure}

To clearly illustrate the role of the strong interaction in the scattering contribution $C_{\rm sca}$, we compare in Fig.~\ref{Fig:the ana without strong} the $S1/2$, $P1/2$, and $P3/2$ partial-wave correlation functions with (opaque) and without (half-transparent) strong interactions. The strong interaction visibly suppresses the $C_{S1/2}$, while its effect on the $C_{P1/2}$ is almost negligible. For the $C_{P3/2}$, the strong interaction substantially distorts the pure Coulomb baseline, producing a prominent enhancement around 140 MeV/$c$ and a dip around 230 MeV/$c$.

\end{document}